%% file: paper.tex
\documentclass[10pt]{llncs}

\usepackage{amsmath}
\usepackage{amssymb}
\usepackage{balance}  
\usepackage{algorithm}
\usepackage{algorithmic}
\usepackage{url}
\usepackage{subfigure}
\usepackage{multirow}
\usepackage{array}
\usepackage[pdftex]{graphicx}
\usepackage{sidecap}
\sidecaptionvpos{figure}{c}

\usepackage[nocompress]{cite}

\usepackage{url}
\urldef{\mailsa}\path|{immanuel.bayer, uwe.nagel, steffen.rendle}@uni-konstanz.de|

\begin{document}

\renewcommand{\algorithmiccomment}[1]{\hfill $\rhd$ #1}
\newcommand{\x}{\mathbf{x}}
\newcommand{\N}{\mathcal{N}}
\newcommand{\argmin}{\mathop{\mathrm{argmin}}}
\newcommand{\argmax}{\mathop{\mathrm{argmax}}}
\newcommand{\sgn}{\mathop{\mathrm{sgn}}}
\newtheorem{Lemma}{Lemma}
\newtheorem{Remark}{Remark}
\newtheorem{Definition}{Definition}
\renewcommand{\O}{\mathcal{O}}
\hyphenation{Mar-kov off-line}
\newcommand{\fullpicwidth}{0.8}
\newcommand{\picwidth}{0.4}

\title{Graph Based Relational Features for Collective Classification}

\author{Immanuel Bayer \and Uwe Nagel\and Steffen Rendle%
\thanks{Current affiliation: Google Inc.}%
}

\institute{
       University of Konstanz\\
       78457 Konstanz, Germany\\
\mailsa\\
}


\maketitle
\input{abstract.tex}
\input{intro.tex}
\input{problem.tex}
\input{method.tex}
\input{related_work.tex}
\input{evaluation.tex}

\section{Conclusion}
We have shown that dependencies between samples can be exploited using relational feature engineering.
Our method allows to combine relational information from various sources with attributes attached to individual samples.
We tested this on standard SRL benchmark datasets, showing that even on network only data our features are competitive to specialized relational learning models.
In addition, our features can outperform them when additional information is available.
Note that in contrast to the SRL methods, our proposal achieves these results without collective inference.
While we restricted our experiments to logistic regression as prediction model, the proposed features could be used as input to any other feature based learning algorithm such as SVM, neural networks or random forests.
Extending the use of relational features to multi relational datasets would be straight forward and a interesting direction for further research.


\subsubsection*{Acknowledgments.}
This work was supported by the DFG under grants Re 3311/2-1 and Br 2158/6-1.


\bibliography{references}
\bibliographystyle{splncs_srt}
\end{document}

%% file: abstract.tex
\begin{abstract}
Statistical Relational Learning (SRL) methods have shown that classification accuracy can be improved by integrating relations between samples.
Techniques such as \textit{iterative classification} or \textit{relaxation labeling} achieve this by propagating information between related samples during the inference process.
When only a few samples are labeled and connections between samples are sparse, \textit{collective inference} methods have shown large improvements over \textit{standard feature-based} ML methods.
However, in contrast to feature based ML, collective inference methods require complex inference procedures and often depend on the strong assumption of label consistency among related samples.
In this paper, we introduce new \textit{relational features} for standard ML methods by extracting information from \textit{direct} and \textit{indirect relations}.
We show empirically on three standard benchmark datasets that our relational features yield results comparable to collective inference methods.
Finally we show that our proposal outperforms these methods when additional information is available.

\end{abstract}


%% file: intro.tex
\section{Introduction}
Statistical relational learning (SRL) methods are used when samples are connected by one or more relation. These relations are helpful in tasks like
scientific article classification where patterns such as ``connected samples have similar labels'' are very predictive.
Feature based ML methods in contrast often assume that samples are independently, identically distributed (iid).
This approach is well established, allows for efficient parameter estimation and simplified prediction but ignores the relational information available in SRL settings.
Recently, a number of methods have been proposed \cite{macskassy2003simple,taskar_probabilistic_2001,lu_link-based_2003,neville_iterative_2000} that significantly improves
over \textit{classical} methods by using joint inference.
Exact joint inference has high runtime complexity which often requires approximate solutions \cite{sen_collective_2008}. These approximated joint inference techniques introduce new difficulties
such as the need for specialized implementations that are expensive to run and difficult to tune \cite{sen_collective_2008}.

In this paper we propose to transfer the relational information into classical features.
This allows a straightforward combination of relational and classical (attribute) information. 
It also renders traditional, feature based ML methods competitive in settings where relational information is available and allows to leverage the large body of classical ML methods and their scalable algorithms.

We use three standard \textit{collective classification} (CC) benchmark datasets to
show that classical ML with relational features are strong competitors for state of the art SRL methods on this task.
Note that on these datasets, collective classification are considered the best performing methods in the current literature \cite{jensen_why_2004, macskassy_classification_2007}.
In particular, we make the following contributions:
\begin{itemize}
    \item We discuss how joint inference could be avoided by extending the sample description with relational features (Section \ref{sec:preserving_iid}).

    \item We extend relational features to indirect relations (Section \ref{sec:rel-features}).
        This is new and crucial to achive high accuracy when only few samples are labeled (Section \ref{sec:relational-features}).
  \item We introduce a new cluster based relational feature (Section \ref{sec:clustering-features}) that provides strong results and is cheap to compute.

  \item We show that our approach improves state of the art collective classification even in network only settings (Section \ref{sec:best_off}).
\end{itemize}


%% file: problem.tex
\section{Problem Setting}\label{sec:problem}
We start by giving the necessary definitions with the traditional setting of samples 
$\mathcal{D}=\{(x_i,y_i)\}_{i=1}^N$ where $y_i$ is the class label and $x_i$ is a feature vector describing sample $i$.
We assume that for the first $u$ samples $(i \in \{1,\ldots,u\})$, the class label $y_i$ is known and for the samples with index $i > u$, the class label is unknown.
Relations among samples are represented by weighted, symmetric adjacency matrices $R_k\in \mathbb{R}^{n\times n}$ and the complete relational information is denoted as $\mathcal{R}=\{R_k\}_{k=1}^K$.
While in the general case, each of the $k$ relations could be complete, i.e.\ provide some similarity between each pair of samples, we explicitly consider the case of sparse, unweighted relations where only a minority of node pairs are connected by an edge.
We start with a statistical argument to motivate the representation of relational information in a way
that is compatible with the iid assumption.

%% file: method.tex
\subsection{Preserving IID}\label{sec:preserving_iid}
Many machine learning algorithms are based on the maximum likelihood principal to learn the optimal value $\theta^*$ for model parameters given a dataset $\mathcal{D}$ (c.f.~\cite{murphy2012machine})
    $$\theta^* := \arg\max_\theta p(\theta | \mathcal{D}) = \arg\max_\theta p(\mathcal{D}|\theta)$$
where
  $l(\theta) := p(\mathcal{D}|\theta)$
is called the likelihood.
A very common assumption in many ML approaches is that the samples in the dataset $D$ are independent and identically distributed (iid).
This assumption simplifies the likelihood to
$$  p(\mathcal{D}|\theta) \stackrel{iid}{\propto} \prod_{i=1}^N p(y_i | x_i, \theta).$$
One of the central arguments of relational learning is that examples are not iid.
In particular for any pair of examples $(x_i, y_i)$ and $(x_j, y_j)$ conditional independence does not hold
$$  p( (x_i, y_i), (x_j, y_j) | \theta) \not \propto p(y_i | x_i, \theta) \, p(y_j | x_j, \theta).$$
Note that in this formulation the relational information $\mathcal{R}$ is completely neglected.
However if $\mathcal{R}$ is used, it can render the probabilities independent
\begin{equation}
\begin{split}
    p( (x_i, y_i)&, (x_j, y_j), \mathcal{R} | \theta)\\ &\propto p(y_i | x_i, \mathcal{R}, \theta) \, p(y_j | x_j, \mathcal{R},\theta). 
\end{split}\label{eq:iidg}
\end{equation}
This formulation is close to standard non-relational ML with iid of samples.

Standard ML algorithms assume that all information about a sample can be encoded in a (usually real-valued) feature vector $x$.
Let us assume that the influence of $\mathcal{R}$ on sample $i$ can be described through a finite number of real valued variables $x^r_i$.
We call $x^r_i$ \emph{relational features}.
To simplify notation, we can define $\tilde{x}_i$ as the extended feature vector of an example $i$ that combines both non-relational features $x_i$ and relational features $x^r_i$.
In total, this allows to rewrite Equation~\ref{eq:iidg} 
\begin{align*}
  p(\mathcal{D}|\theta) \stackrel{iid}{\propto} \prod_{i=1}^N p(y_i | \tilde{x}_i, \theta). \label{eq:iidrel}
\end{align*}
Note that due to the relational information in $\tilde{x}_i$, the iid assumption can be preserved.
In the remainder of this section, we discuss several ways to generate relational features.

\subsection{Graph Based Relational Features}\label{sec:rel-features}
Samples can be linked through multiple relations each of which can be described as a graph.
Representing each relation by an independent set of features allows us to integrate an arbitrary number of relations per problem into a standard feature matrix.
All of the proposed features have in common that the encoded relational information does not only consist of direct relations but in addition captures indirect relations which we found to be the key to their performance.

\subsubsection*{Neighbor Ids}\label{sec:neighbor-features}
\begin{SCfigure}
\centering
    \includegraphics[width=0.48\textwidth]{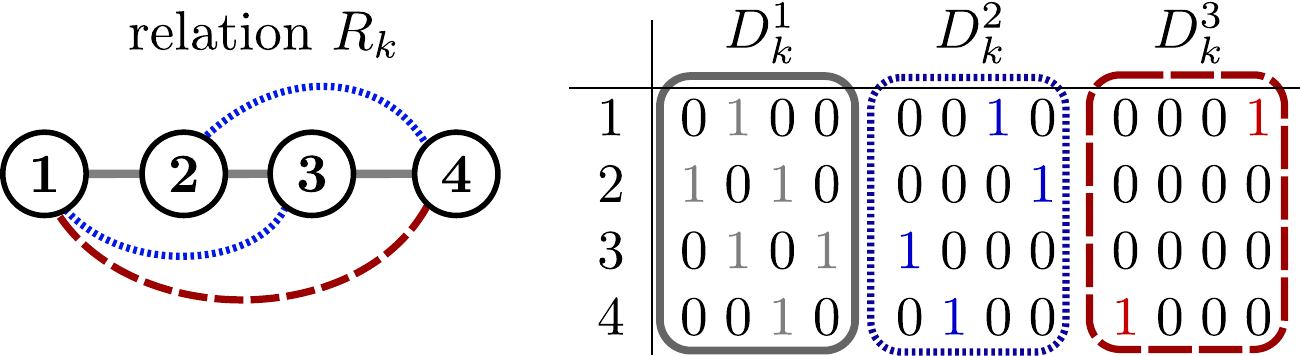}
\caption{Feature matrix for the node neighborhoods of a relation. The edges of the original relation are depicted as gray lines, while connections in distance two and three are shown as blue/short dashed and red/long dashed curves.}\label{fig:adjacency-encode}
\end{SCfigure}

Encoding the direct neighbors of a sample $i$ in relation $R_k$ can be achieved by treating each sample as a categorical variable
which is true when samples are connected and false otherwise \cite{bernstein_relational_2003, perlich2003aggregation, perlich_distribution-based_2006}.
As this information is very local and yields limited information about $i$'s position in $R_k$, we extend it
by additionally including indirect neighbors at various distances to $i$. 
Distance refers to the number of edges  $d_k(i,j)$ on a shortest path connecting $i$ and $j$ in $R_k$.
In particular, for a (small) set of distances $d\geq 1$ we describe each relation $R_k$ by
the distance-$d$ neighborhood matrices $D^d_k\in \{0,1\}^{n\times n}$ with $\left(D^d_k\right)_{i,j}=1$, if $d_{k}(i,j)=d$ and 0 otherwise.
We illustrate the derivation of this feature in Figure~\ref{fig:adjacency-encode}.


\subsubsection*{Aggregated Neighbor Attributes}
\begin{SCfigure}
\centering
    \includegraphics[width=0.48\textwidth]{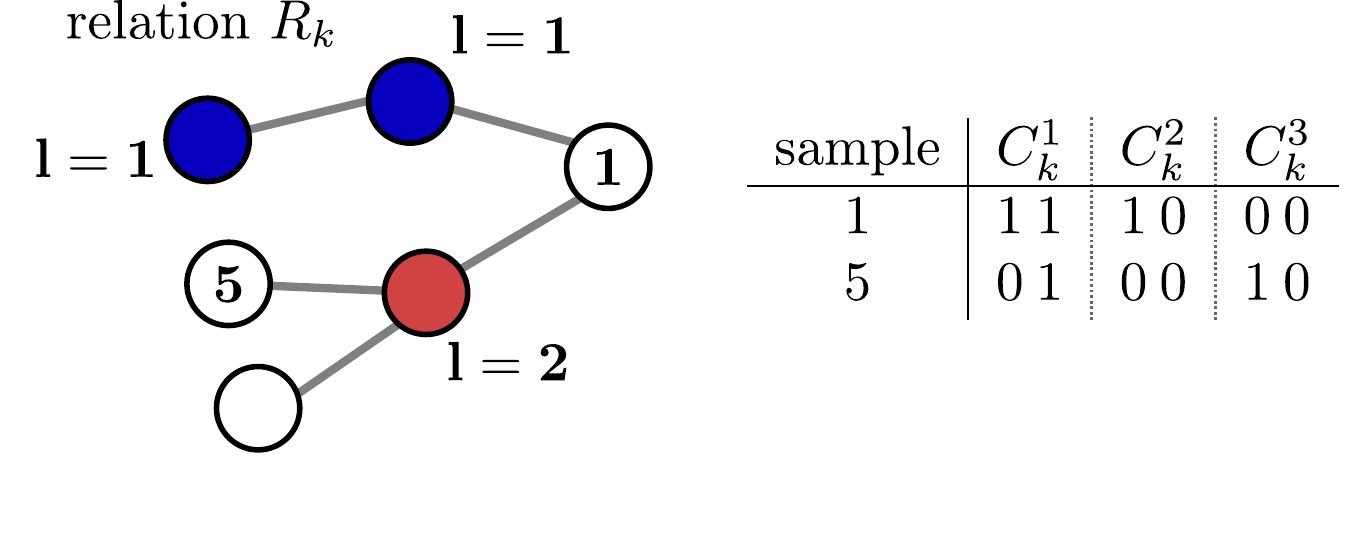}
\caption{Aggregation for attribute counts on distances one, two and three in the relation $R_k$. Labels are given in text and as color (1-blue, 2-red).
Sample 1 has a single node of each label as direct neighbors which is reflected in the first two columns($C_k^1$) of its row, while the next two columns ($C_k^2$) encode the single 2-labeled node in distance 2.
Note that unlabeled nodes (white) are ignored in the features.
}\label{fig:aggregation-encode}
\end{SCfigure}

Relational position can also be described by individual features of direct and indirect neighbors \cite{neville_learning_2003,neville_simple_2003}.
As before we extend the idea by calculating individual features at various distances.
For a categorical attribute with categories $1,\ldots,c$, we define an $n\times c$ matrix $L$ with $L_{i,j}=1$ if $x_i$ is labeled as $j$ and $L_{i,j}=0$ otherwise.
Then the count matrix $C_k^d=:D_k^dL$ can be derived as the projection of the corresponding neighborhood matrix to the label matrix $L$ such that
$(C_k^d)_{i,j}$ yields the number of category $j$ nodes in distance $d$ of sample $i$.
We denote this feature by \emph{neighbor class counts} (NCC) and provide an illustration in Figure~\ref{fig:aggregation-encode}.
An additional row normalization yields a probability matrix for the class labels in distance $d$, which we will denote by \emph{neighbor class probabilities} (NCP).

\subsubsection*{Random Walk Similarity}\label{sec:rwr-features}
While the features described above are based strictly on shortest paths, random walks with restart (rwr) incorporate a different notion of connectivity.
They have been proposed as a similarity measure in the context of auto-captioning of images~\cite{pan_automatic_2004}.

The similarity between two nodes is measured as the probability of a random walk connecting them, i.e.\ the probability of the random walk process visiting one node when started from the other.
To control for locality this includes a restart probability: in each step the walk will jump back to the starting point with probability $r$.
This can be modeled as
\begin{equation}
    p_i = (1-r)W p_i + re_i\;
    \label{eq:rwr}
\end{equation}
where the column vector $p_i$ of matrix $P$ describes the steady-state probability distribution over all samples for walks starting at $i$. 
$W$ is the matrix encoding transition probabilities, i.e.\ a $L_1$ row normalized version of $R_k$, $e_i$ is the vector of zeros and unit value at position $i$ and the parameter $r$ is the restart probability.
$P$ can be determined as the solution of a linear system or approximated efficiently~\cite{tong_fast_2006}, leaving $r$ as free parameter.
We derive $\hat{P}$ from $P$ by column wise $L_2$ normalization and use analogous to the neighbor features row $i$ of $\hat{P}$ as features for sample $i$.

\begin{SCfigure}
\centering
    \includegraphics[width=0.48\textwidth]{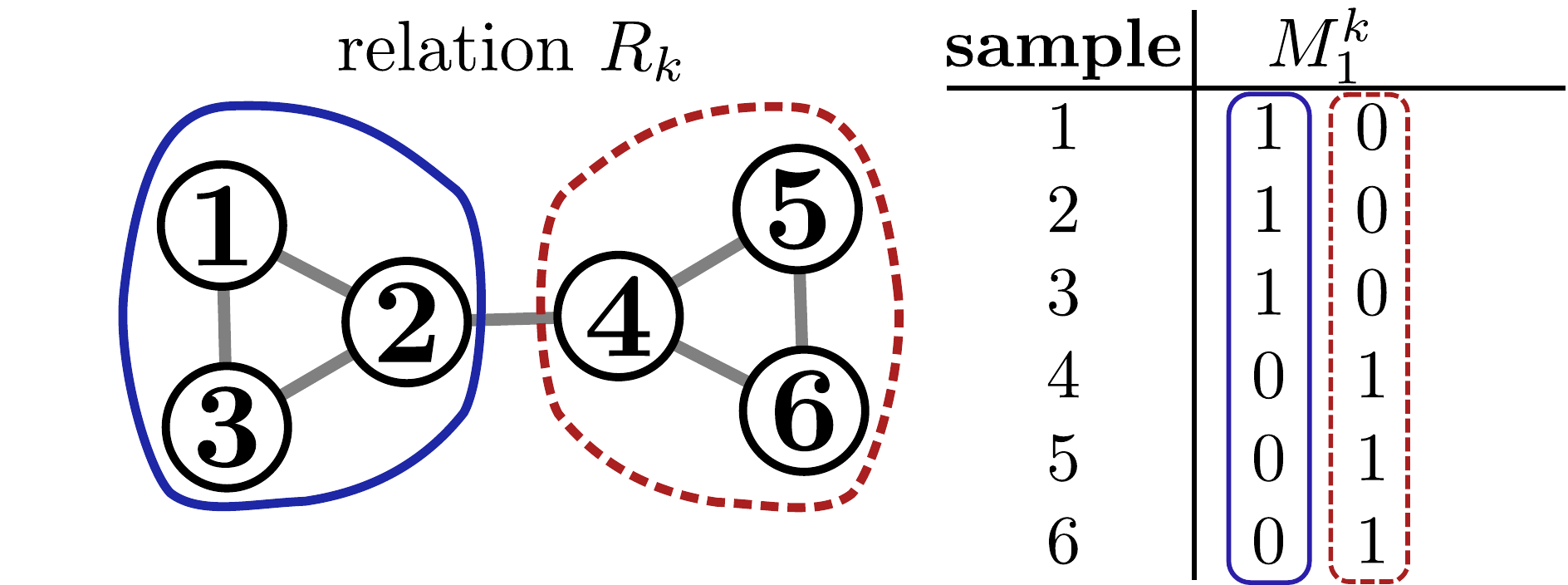}
\caption{An example for encoding a clustering of a relation as a feature vector. 
Memberships of samples in clusters are represented as binary features.}\label{fig:cluster-encode}
\end{SCfigure}
\subsubsection*{Clustering Memberships}\label{sec:clustering-features}
Clustering methods can be used to identify groups of similar samples.
Clustering features encode this information by representing this group membership.
Given a clustering of the graph representing relation $R_k$ into $c$ clusters, we obtain an $n\times c$ feature matrix $M_k^c$ with $(M_k^c)_{i,m}=1$ if sample $i$ belongs to cluster $m$ and zero otherwise.
Since a single clustering yields limited information about the dense groups in the graph, we create features for various clusterings, i.e.\ different $c$.
We limit $c$ to $c=2^j$ subject to $2\leq c\leq n$ which limits the number of clusterings to $\lfloor\log_2(n)\rfloor$ while also providing a wide range of cluster sizes.
This results in $O(n)$ features per relation that are very sparse with only $O(\log(n))$ non-zero features for per sample.
The clusters can be calculated with negligible runtime using the METIS clustering framework\footnote{We used the implementation available at \url{http://glaros.dtc.umn.edu/gkhome/views/metis} with default parameters.} \cite{karypis1998fast}. 
Note that the dense subgroups identified in the clusterings can be directly related to the homophily assumption often exploited in relational learning.


%% file: related_work.tex
\section{Related Work}\label{sec:rel-work}
We build on two main categories of related work. The first, in Section \ref{sec:relational_features} uses features derived from
the network structure to improve iid based inference.
The second, discussed in Section \ref{sec:collective_inference}
is work that views collective classification as a joint inference problem,
simultaneously inferring the class label on every instance.
The challenges specific to problems with few labeled data points have received special attention
\cite{macskassy_classification_2007, saar-tsechansky_handling_2007, gallagher_using_2008, gallagher_leveraging_2010}
and helped us to understand the importance of indirect relations.

\subsection{Relational Features}\label{sec:relational_features}
Relational features can be combined with collective inference or directly used with standard ML methods
as we argue in Section \ref{sec:preserving_iid}. Models such as Relational Probabilistic Trees~\cite{neville_learning_2003},
Relational Bayes Classifier~\cite{neville_simple_2003} and the Link Based Classifier (LBC, \cite{lu_link-based_2003}) concentrate primarily on the aggregation of
attributes of connected samples. Others use rows from the (weighted) adjacency matrix as basis for feature construction
~\cite{bernstein_relational_2003,perlich_distribution-based_2006,perlich2003aggregation}.
We were especially inspired from the suggestion to extend the neighborhood of samples with few neighbors with distance-two neighborhoods~\cite{preisach_relational_2006}
or ghost edges \cite{gallagher_using_2008}. In contrast to previous work we keep the information
from various neighborhood distances separated and introduce the concept of multiple indirect relations.

\subsection{Collective Inference}\label{sec:collective_inference}
Full relational models such as Markov Logic Networks (MLN, \cite{richardson_markov_2006}) or the
Probabilistic Relational Model \cite{taskar_probabilistic_2001} can be used for CC \cite{crane_investigating_2012}.
We refer to Sen et al. \cite{sen_collective_2008} for a comprehensive overview of collective inference based CC algorithms.
Their strength in high label autocorrelation settings and the problem of error propagation has been examined \cite{jensen_why_2004, xiang2011understanding} and improved inference
schemes have been proposed \cite{mcdowell2009cautious}.
Recently, stacking of non relational model has been introduced \cite{kou2007stacked} as a fast approximation of
Relational Dependency Networks \cite{neville_collective_2003}.


%% file: evaluation.tex
\section{Evaluation}
In our experiments\footnote{Our code is available from https://github.com/ibayer/PAKDD2015} we investigate the following three questions:
\begin{enumerate}
    \item Are classical ML methods with relational features competitive to MLN and Collective Classification approaches.
    \item What are the main ingredients that make relational features effective.
    \item Does the combination of relational and attribute information improve results?
\end{enumerate}

\subsection[Experimental Setup]{Experimental Setup
}\label{sec:evaluation}

\textbf{Datasets}
We use three standard benchmark SRL datasets.
The \emph{Cora} and \emph{CiteSeer} scientific paper collections have been used in different
versions, we chose the versions\footnote{http://linqs.cs.umd.edu/projects/projects/lbc/index.html} presented in~\cite{sen_collective_2008} and the \emph{IMDb} dataset\footnote{http://netkit-srl.sourceforge.net/data.html}.
Both, Cora and CiteSeer include text features in form of bag of words (bow) representations.
We give some statistics of these datasets in Table~\ref{table:datasets}.\\
\begin{table}[t]
  \begin{center}
    \begin{tabular}{l*{5}{c}r}
      & Nodes &  Links & Classes & $|\text{Dictionary}|$ & avg. Degree \\
      \hline
      Cora          & 2708 & 5278  & 7 & 1433  & 3.8981   \\
      CiteSeer      & 3312 & 4660  & 6 & 3703  & 2.8140   \\
      IMDb (all)    & 1377 & 46124 & 2 & -     & 66.9920  \\
    \end{tabular}
  \end{center}
\caption{Summary statistics of the datasets.}
    \label{table:datasets}
\end{table}
\textbf{Benchmark Models} As baseline models we use the well established relational learning methods wvRN~\cite{macskassy2003simple,macskassy_improving_2007}, nLB~\cite{lu_link-based_2003} and MLN~\cite{crane_investigating_2012}.
We chose \emph{relaxation labeling}~\cite{chakrabarti_enhanced_1998} as the collective inference method for wvRN and nLB as it has been shown to outperform Gibbs sampling and iterative classification on our datasets~\cite{macskassy_classification_2007}.
For MLN we used the rules \verb|HasWord(p,+w)=>Topic(p,+t)| and \verb|Topic(p,t)^Link(p,p')=>Topic(p',t)| together with discriminative weight learning and MC-SAT inference as recommended for Cora and Citeseer in a previous study \cite{crane_investigating_2012}.
\\ \textbf{Measures \& Protocol}
We follow~\cite{macskassy_classification_2007} and remove samples that are not connected to any other sample (singletons) in all experiments.
Each experiment is repeated 10 times with class-balanced splits into ratios of $0.1, 0.2, \dots, 0.9$ for the train-test set (shown as percentage in the figures). 
The MLN experiments are only repeated 3 times due to their extremely high evaluation time.
We calculate multi-class accuracies by micro averaging and plot them on the y-axis of each figure in this section. 
We used the netkit-srl framework~\cite{macskassy_classification_2007} in version (1.4.0)\footnote{http://netkit-srl.sourceforge.net/} to evaluate the wvRN and nLB classifiers and
 the Alchemy 2.0 framework\footnote{https://code.google.com/p/alchemy-2/} to evaluate the MLN model.
The graph clusterings were obtained using METIS \cite{karypis1998fast} version 5.1.0.
Relational features are learned with an $L_2$ penalized logistic regression model\footnote{We use a one-vs-all scheme for multiclass classification.} included in scikit-learn~\cite{scikit-learn} version 0.14.01.
The penalty hyperparameter $C$ is optimized via grid search over $\{0.001,0.01,0.1,1,10,100,1000\}$ on the training set.

\subsection{Comparing Relational Features to SRL}\label{sec:best_off}

This section examines whether feature based relational models (without collective inference) are able to compete with the prediction quality of specialized relational learning methods. 
Consequently, the benchmark is a task where only relational data is available.
In the first experiment, we compare wvRN and nLB with two logistic regression models that use only our relational features.
The first relational feature model (rwr) is based on a random walk.
The second model uses both, neighborhood and aggregated (NCP) features with distances 1,2,3.
We exclude distances higher than three, since almost every node can be reached over three edges from every other node and therefore further distances do not provide additional information.

\begin{figure*}[h]
    \centering
    \includegraphics[width=\fullpicwidth\textwidth]{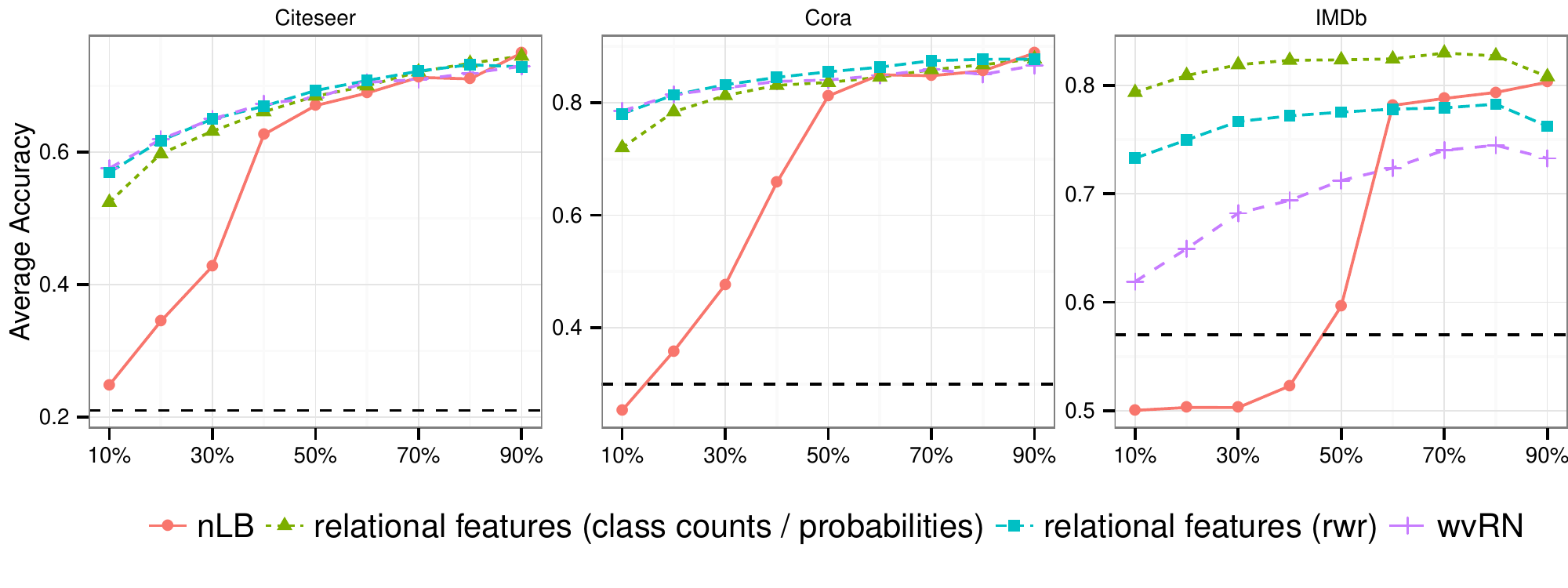}
    \caption{Comparison between two SRL methods (nLB, wvRN) with relaxation labeling and two relational
    feature based models on network only data. The dashed black lines indicate base accuracy.
}
 \label{fig:best_off}
\end{figure*}

Figure~\ref{fig:best_off} illustrates two problems of SRL models: 
(i) nLB performs poorly\footnote{This has been attributed to a lack of training data \cite{macskassy_classification_2007}.} when labels are sparse and
(ii) wvRN is sensitive to violations of its built in assumptions -- i.e. if label consistency among neighbors is not met, as with the IMDb dataset.

The relational feature based models show a very consistent performance not much affected by the number of labeled samples. 
The results of the neighbor and NCP feature combination on IMDb illustrate the flexibility of relational features.


\subsection{Engineering Relational Features}\label{sec:relational-features}
\begin{figure*}[h]
    \centering
    \includegraphics[width=\fullpicwidth\textwidth]{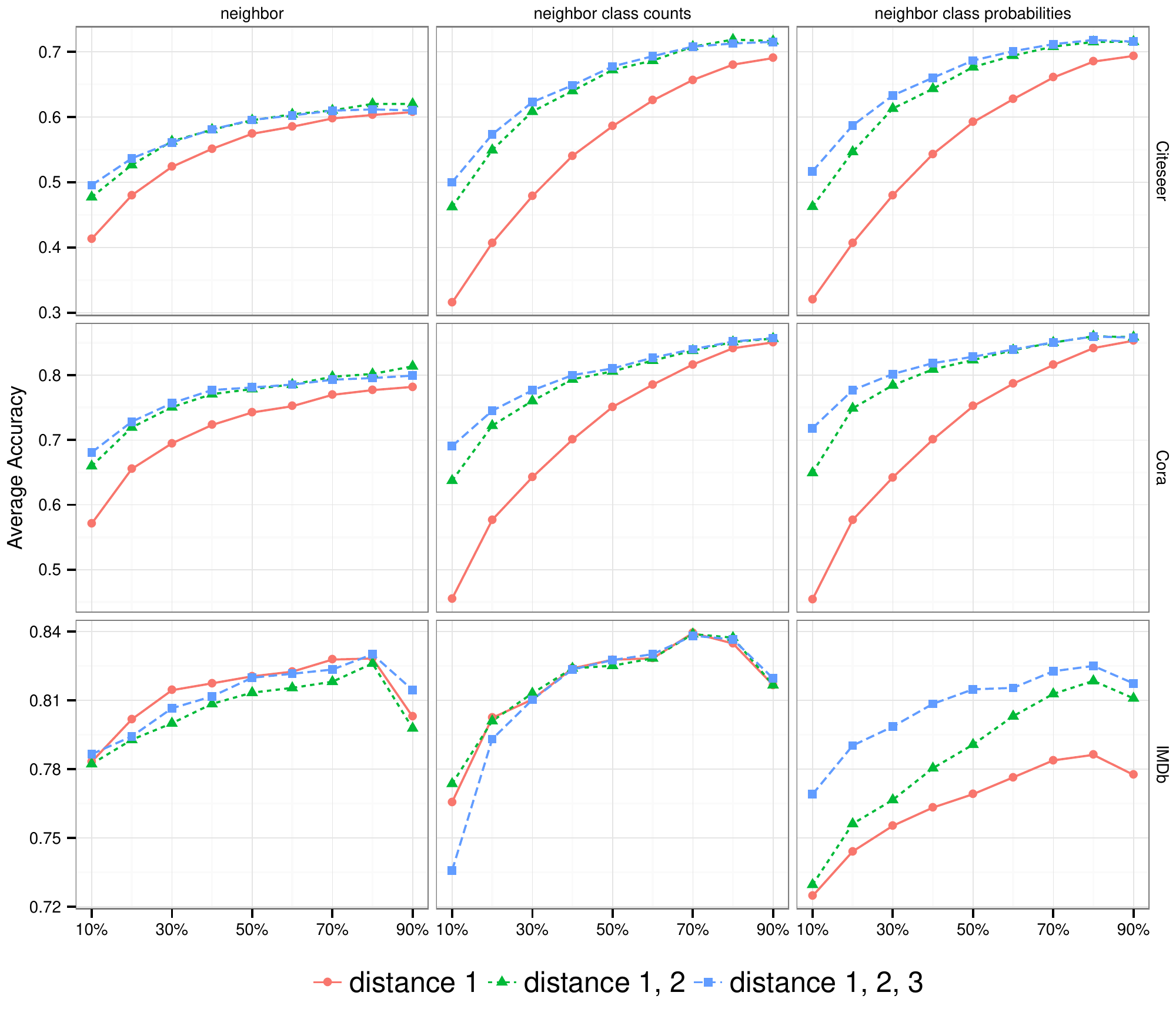}
    \caption{Influence of the distance parameter for label dependent and independent
    relational features. Including information from indirect neighbors improves
    results especially if few samples are labeled.}
 \label{fig:depth}
\end{figure*}
\begin{figure*}[h]
    \centering
    \includegraphics[width=\fullpicwidth\textwidth]{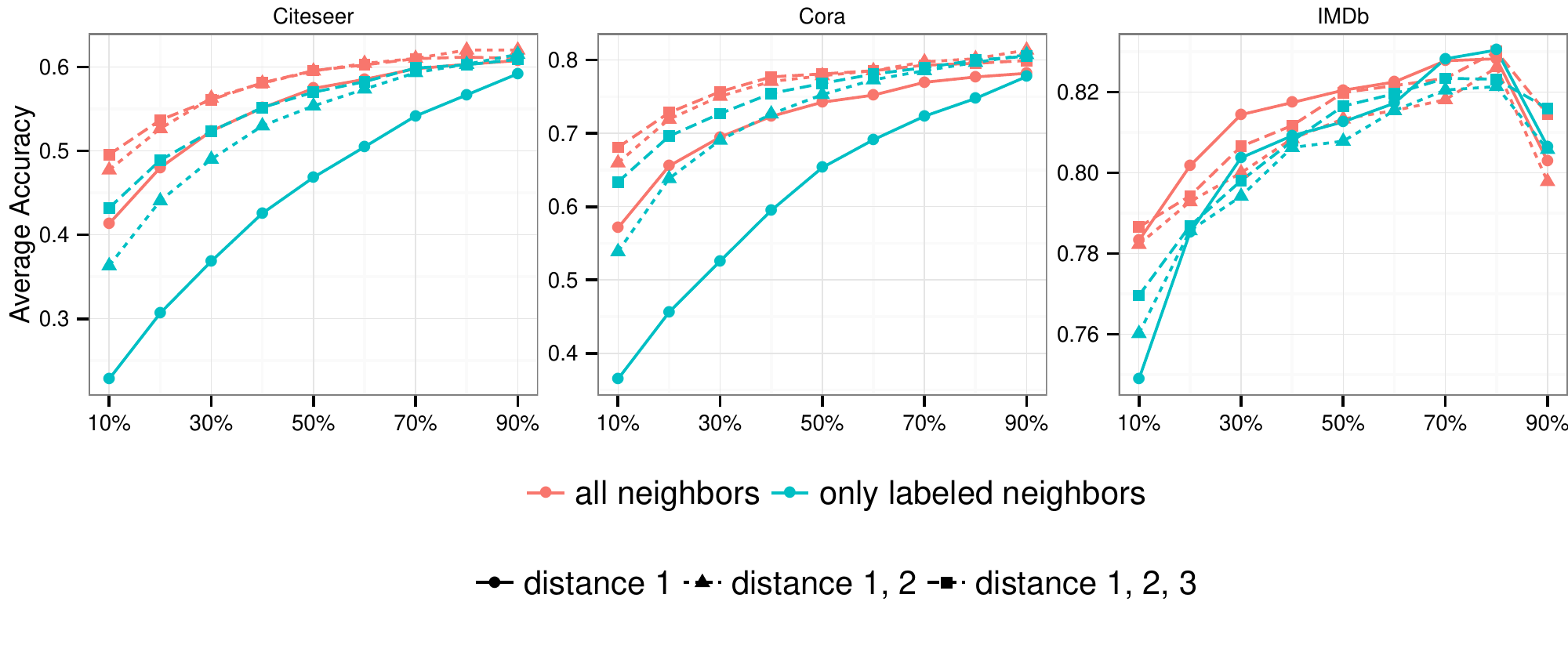}
    \caption{Comparison of using only labeled vs all neighbors to construct
        relational neighbor features (NCC). Including unlabeled neighbors  improves the results
        in all settings.}
 \label{fig:ids_labeled}
\end{figure*}
\begin{figure*}[h]
    \centering
    \includegraphics[width=\fullpicwidth\textwidth]{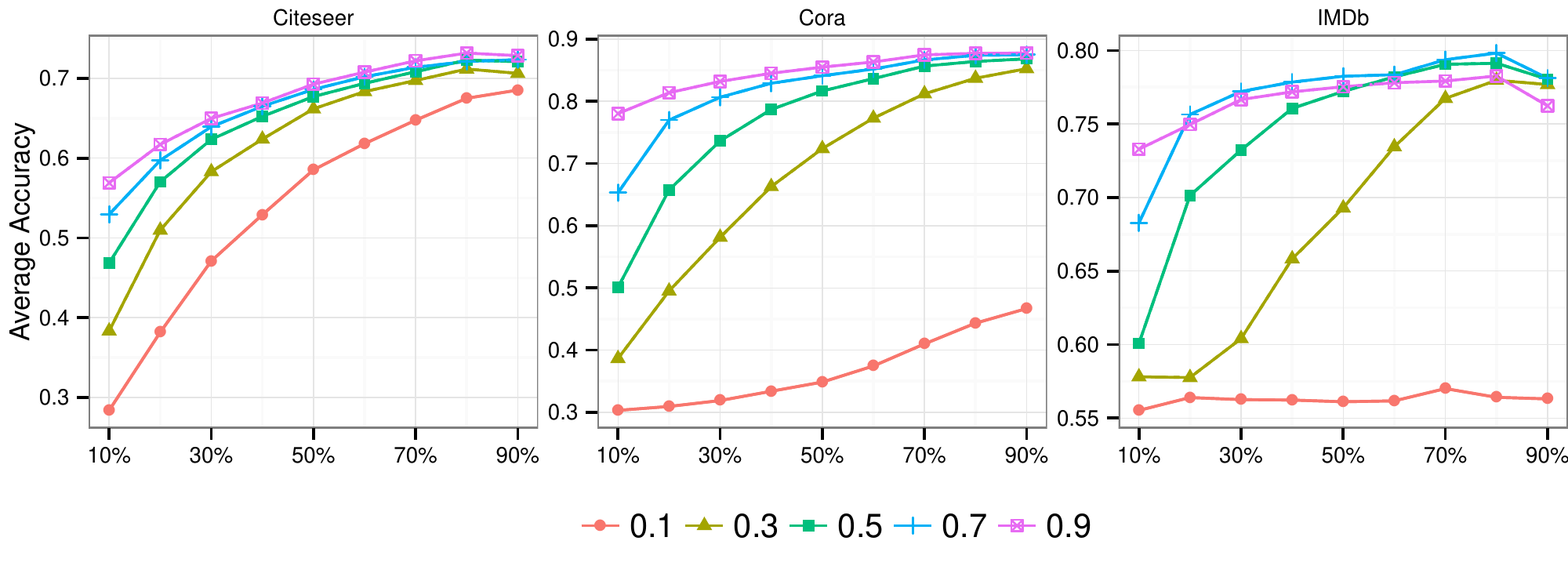}
    \caption{Influence of various restart values on the prediction accuracy. 
While as a general trend larger values of $r$ (c.f.\ Equation~\ref{eq:rwr}) lead to better results, the value $r=0.9$ yields consistently good results.}
 \label{fig:rwr}
\end{figure*}

\begin{figure*}[h]
    \centering
    \includegraphics[width=\fullpicwidth\textwidth]{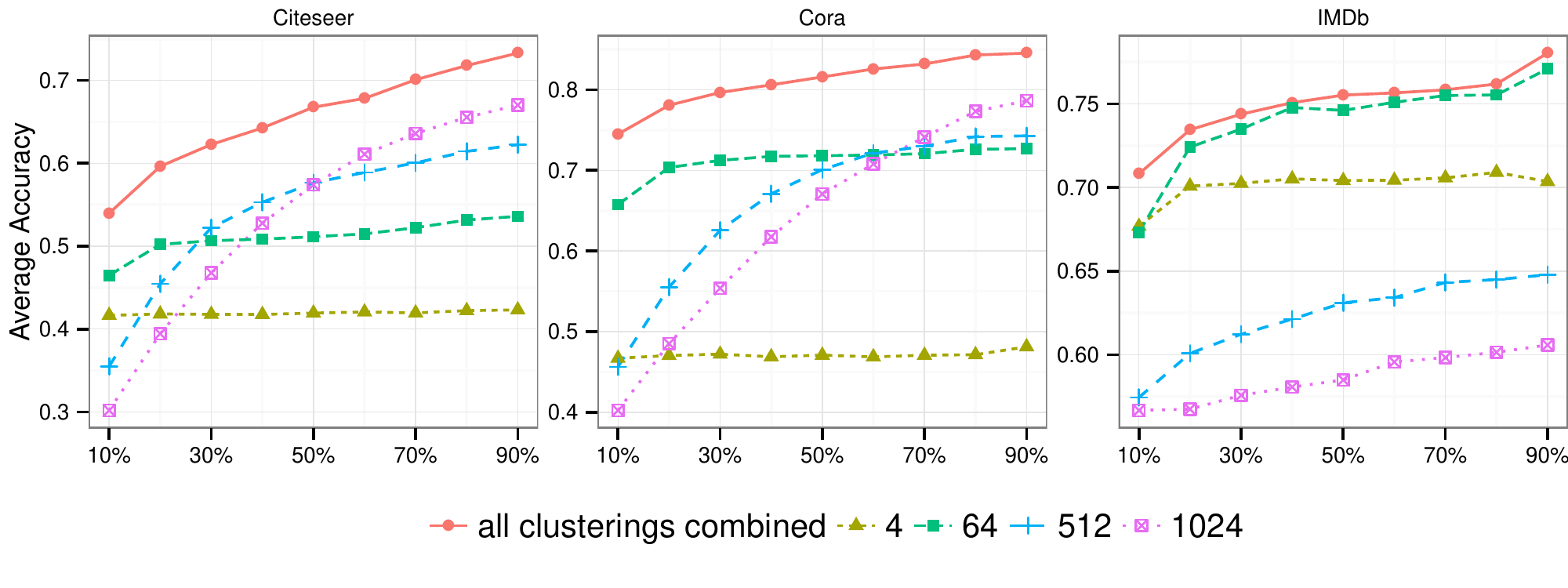}
    \caption{Examination of various clusterings of the dependency network and their usage as relational feature. 
The optimal number of clusters strongly depends on the dataset and ratio of labeled samples.
The combination of all clusterings, however, is consistently superior to individual clusterings.}
 \label{fig:cluster}
\end{figure*}

We now examine the different relational features and the influence of their parameters.
Two questions will be addressed:
(i) \textit{How important are indirect relations?} 
(ii) \textit{Which of the proposed relational features lead to the best results?}
All relational features that we consider can incorporate information about indirect neighbors. 
Each method has parameters that adjusts the locality of the resulting features.
We first examine the effect of including indirect neighbors (Figure~\ref{fig:depth}) and
the importance of unlabeled neighbors (Figure \ref{fig:ids_labeled}) for relational neighbor features.
The influence of the restart parameter $r$ (c.f. Equation~\ref{eq:rwr}) for the rwr features can be seen in Figure~\ref{fig:rwr} and an informative subset of results
for various numbers of clusters is shown in Figure~\ref{fig:cluster}.
The results suggest that the inclusion of indirect neighbors in the relational features is beneficial independently of whether they are used directly or for aggregation. 
Figure~\ref{fig:ids_labeled} shows that unlabeled neighbors contribute significantly to the overall performance.
Together this answers our first question: unlabeled samples and indirect neighborhood relations are essential ingredients for relational features.

\subsection{Combining Relational and Local Information}\label{sec:rel-loc-comb}
\begin{SCfigure}
\centering
    \includegraphics[width=0.48\textwidth]{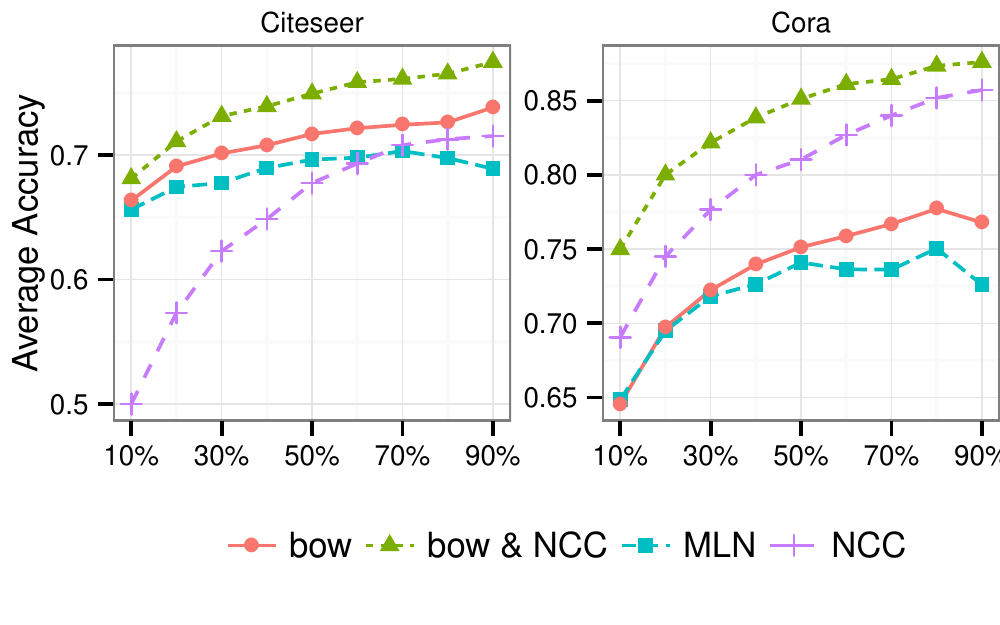}
    \caption{Comparison of attribute only (bow), network only (NCC), MLN and our combination of relational features with local attributes.
  }
 \label{fig:bow}
\end{SCfigure}
In the following, we examine the effect of adding local attributes.
Figure~\ref{fig:bow} shows results with neighborhood count features (NCC) of distances 1,2,3.
Interestingly, the bag of words model performs better than network only models on Citeseer but worse on Cora.
Combining relational and local attributes on the other hand, improves results in both cases.
The figure further shows that our features outperform MLN on both datasets.
In summary, our experiments suggest that the combination of relational features and attributes is beneficial even with a simple model such as logistic regression.
\subsection{Discussion}

Our experiments indicate that relational feature based models compare well to specialized relational learners even in network only and sparse labeling settings. 
This has been verified on three standard SRL benchmark datasets and with three state of the art SRL methods for comparison.
The inclusion of indirect neighbors has proven extremely important, especially in sparse label settings.
We have further shown that the combination of relational features and local attributes is both straightforward and has the potential to improve considerably over both, feature only and network only models.

Note, that our relational features can lead to very high dimensional representations.
Such feature spaces are, however, common in recommender systems, click-through rate prediction and websearch where regularized logistic regression has been shown to be very effective \cite{graepel2010web, liu2009large}.
In addition, we use a standard implementation of logistic regression and can consequently
employ scalable versions that can be trained with billions of samples of high dimensions \cite{mukherjee2013parallel, rendle2013scaling}.
We are further not committed to the logistic regression model as our features could be used as input for arbitrary vector space models.


